
\input phyzzx

\def\IR{{\hbox{{\rm I}\kern-.2em\hbox{\rm R}}}}
\def\IB{{\hbox{{\rm I}\kern-.2em\hbox{\rm B}}}}
\def\IN{{\hbox{{\rm I}\kern-.2em\hbox{\rm N}}}}
\def\IC{{\ \hbox{{\rm I}\kern-.6em\hbox{\bf C}}}}

\def\IZ{{\hbox{{\rm Z}\kern-.4em\hbox{\rm Z}}}}
\def\to{\rightarrow}
\def\d{{\rm d}}
\def\underarrow#1{\vbox{\ialign{##\crcr$\hfil\displaystyle
{#1}\hfil$\crcr\noalign{\kern1pt
\nointerlineskip}$\longrightarrow$\crcr}}}
%
\def\d{{\rm d}}
\def\ltorder{\mathrel{\raise.3ex\hbox{$<$}\mkern-14mu
             \lower0.6ex\hbox{$\sim$}}}
\def\lesssim{\mathrel{\raise.3ex\hbox{$<$}\mkern-14mu
             \lower0.6ex\hbox{$\sim$}}}


\input phyzzx
\overfullrule=0pt
\tolerance=5000
\overfullrule=0pt
\twelvepoint

\twelvepoint
\pubnum{IASSNS-HEP-93/10}
\date{March, 1993}
\titlepage
\title{ON THE LANDAU-GINZBURG DESCRIPTION OF $N=2$ MINIMAL MODELS}
\vglue-.25in
\author{Edward Witten
\foot{Research supported in part by NSF Grant
PHY92-45317.}}
\medskip
\address{School of Natural Sciences
\break Institute for Advanced Study
\break Olden Lane
\break Princeton, NJ 08540}
\bigskip
\abstract{The conjecture that $N=2$ minimal models in two dimensions
are critical points of a super-renormalizable Landau-Ginzburg model
can be tested by computing the path integral of the Landau-Ginzburg
model with certain twisted boundary conditions.  This leads to simple
expressions for certain characters of the $N=2$ models which can be verified
at least at low levels.
An $N=2$ superconformal algebra can in fact be found directly in the
{\it noncritical} Landau-Ginzburg system, giving further support
for the conjecture.}
\endpage

\chapter{Introduction}

$N=2$ supersymmetry in two dimensions has several distinguished
classes of simple realizations.  In the present
paper, we will focus on two of these, and the conjectured relationship
between them.

\REF\divec{P. DiVecchia, J. L. Peterson, and H. B. Zheng,
``$N=2$ Superconformal Theory In Two Dimensions,'' Phys. Lett. {\bf 162B}
(1985) 327}
\REF\newv{P. DiVecchia, J. L. Petersen and M. Yu, ``On The Unitary
Representations of $N=2$ Superconformal Theory,'' Phys. Lett.
{\bf B172} (1986) 211; P. DiVecchia, J. L. Petersen, M. Yu,
and H. B. Zheng, ``Explicit Construction Of Unitary Representations
Of The $N=2$ Superconformal Algebra,'' Phys. Lett. {\bf B174} (1986) 280.}
\REF\friedan{W. Boucher, D. Friedan, and A. Kent, ``Determinant Formulae
and Unitarity For the $N=2$ Superconformal Algebra In Two
Dimensions, Or Exact Results On String Compactification,''
Phys. Let. {\bf B172} (1986) 316.}
\REF\cappelli{A. Cappelli, C. Itzykson, and J.-B. Zuber, ``Modular Invariant
Partition Functions In Two Dimensions,'' Nucl. Phys. {\bf B280 [FS 18]}
(1987) 445; A. Cappelli, ``Modular Invariant Partition Functions Of
Superconformal Theories,'' Phys. Lett. {\bf B185} (1987) 82.}
\REF\gz{D. Gepner and Z. Qiu, ``Modular Invariant Partition Functions
For Parafermionic Field Theories,'' Nucl. Phys. {\bf B285 [FS 19]}
(1987) 423.}
\REF\gepner{D. Gepner, ``Space-Time Supersymmetry In Compactified String
Theory And Superconformal Models,'' Nucl. Phys. {\bf B296} (1988) 757.}
\REF\bagger{J. Wess and J. Bagger, {\it Supersymmetry And Supergravity},
Princeton University Press (second edition, 1992).}
\REF\lastwitten{E. Witten, ``Phases Of $N=2$ Models In Two Dimensions,''
IASSNS-HEP-93/3, to appear in Nucl. Phys. B.}
On the one hand, there is a discrete series of representations of the
$N=2$ superconformal algebra
[\divec --\friedan]
with $\widehat c<1$, in fact with
$$\widehat c = 1-{2\over k+2},~~k=1,2,3,\dots     \eqn\ambo$$
Based on these representations, one can construct families of quantum
field theories known as $N=2$ minimal models.  Actually, these
models have an ${\bf A}-{\bf D}-{\bf E}$
classification [\cappelli--\gepner]; in this paper we consider
only the simplest ${\bf A}$ series.

On the other hand, there are super-renormalizable
Landau-Ginzburg models, constructed from chiral superfields.  Such
a superfield has an expansion
$$\Phi(x,\theta)= \phi(y)+\sqrt 2\theta^\alpha\psi_\alpha(y)
+\theta^\alpha\theta_\alpha F(y),\eqn\bambo$$
where $x,\theta$ are coordinates of two dimensional $N=2$ superspace,
and $y^m=x^m+i\theta^\alpha\sigma^m_{\alpha\dot\alpha}\overline\theta{}
^{\dot\alpha}$; all conventions are as in [\bagger,\lastwitten].
In the simplest case of interest, there is only one chiral superfield,
and the superspace Lagrangian is
$$L=\int\d^2x\,\d^4\theta\,\overline\Phi\Phi
-\int\d^2x\,\,\d^2\theta\,\,{\Phi^{k+2}\over k+2}-\int\d^2x\,\,\d^2\overline
\theta{\overline \Phi^{k+2}\over k+2}. \eqn\cambo$$
The function $\Phi^{k+2}/(k+2)$ that appears here is called the superpotential;
it is a homogeneous function, a fact that ensures the existence
of $R$-symmetries that will play a crucial role later.

\REF\gates{S. J. Gates, Jr., M. T. Grisaru, M. Rocek, and W. Siegel,
{\it Superspace, or One Thousand And One Lessons In Supersymmetry}
(Benjamin-Cummings, 1983).}
\REF\zam{A. Zamolodchikov, ``Conformal Symmetry And Multicritical Points
In Two Dimensional Quantum Field Theory,''
Sov. J. Nucl. Phys. {\bf 44} (1986) 529.}
\REF\kastor{D. Kastor, E. Martinec, and S. Shenker,
``RG Flow In $N=1$ Discrete Series,'' Nucl. Phys. {\bf B316} (1989) 590.}
\REF\martinec{E. Martinec, ``Criticality, Catastrophes, and
Compactifications,''
in {\it Physics And Mathematics Of Strings}, ed. L. Brink, D. Friedan,
and A. M. Polyakov (World Scientific, 1990).}
\REF\warnervafa{C. Vafa and N. Warner, ``Catastrophes And The Classification
Of Conformal Field Theories,'' Phys. Lett. {\bf 218B} (1989) 51.}
\REF\lerche{W. Lerche, C. Vafa, and N. Warner, ``Chiral Rings In
$N=2$ Superconformal Theory,'' Nucl. Phys. {\bf B324} (1989) 427.}
\REF\howe{P. S. Howe and P. C. West, ``Chiral Correlators In Landau-Ginzburg
Models And $N=2$ Superconformal Models,'' Phys. Lett. {\bf B227} (1989) 397,
``Fixed Points In Multi-Field Landau-Ginzburg Models,'' Phys. Lett.
{\bf B244} (1990) 270.}
\REF\ccec{S. Cecotti, L. Girardello, and A. Pasquinucci, ``Non-perturbative
Aspects And Exact Results For The $N=2$ Landau-Ginzburg Models,''
Nucl. Phys. {\bf B238} (1989) 701, ``Singularity Theory And $N=2$
Supersymmetry,'' Int. J. Mod. Phys. {\bf A6} (1991) 2427.}
\REF\cec{S. Cecotti, ``$N=2$ Landau-Ginzburg vs. Calabi-Yau Sigma Models:
Nonperturbative Aspects,'' Int. J. Mod. Phys. {\bf A6} (1991) 1749.}
\REF\fen{P. Fendley and K. Intriligator, ``Scattering And Thermodynamics
Of Fractionally Charged Supersymmetric Solitons,'' Nucl. Phys. {\bf B372}
(1992) 533, ``Scattering And Thermodynamics In Integrable $N=2$ Theories,''
Nucl. Phys. {\bf B380} (1992) 385; S. Cecotti, P. Fendley, K. Intriligator,
and C. Vafa, ``A New Supersymmetric Index,'' Nucl. Phys. {\bf B386}
(1992) 405.}
The non-renormalization theorems for the superpotential of $N=2$ models
(see for instance [\gates, p. 358])
strongly suggest that the $\Phi $ superfield is exactly massless in the
Landau-Ginzburg model with this superpotential.  It has been
conjectured that, in the infrared limit, the interactions of this massless
superfield are precisely governed by the ${\bf A}$ series
of $N=2$ minimal models at level $k$.  This conjecture, which followed
a somewhat analogous discussion for $N=0$ by Zamolodchikov
[\zam] and was motivated in part by
results of Gepner [\gepner], was formulated and tested in [\kastor--\fen].
The goal of the present paper is to exploit and shed light on this
conjecture.

Any observable that is effectively computable in both the Landau-Ginzburg
and minimal models can of course serve as the basis for a test of their
conjectured relation.  However, generic correlation functions of the
Landau-Ginzburg model are not effectively computable, especially in the
infrared where the comparison must be made.
In practice, evidence for the conjecture has come largely
from heuristic renormalization group arguments, comparison of renormalization
group flows for large $k$ [\kastor], a $2+\epsilon$ expansion [\howe] (with
$\epsilon\sim 1/k$), a heuristic computation of the central charge
[\warnervafa], comparison of chiral rings and other properties of
the chiral primary fields [\lerche,\ccec,\cec], and studies of
properties such as soliton scattering and spectrum in integrable
deformations [\fen].

\REF\warner{
A. Schellekens and N. Warner, ``Anomalies And Modular Invariance In
String Theory,'' Phys. Lett. {\bf 177B} (1986) 317, ``Anomaly Cancellation And
Self-Dual Lattices,'' Phys. Lett. {\bf 181B} (1986) 339; K. Pilch,
A. Schellekens, and N. Warner, ``Anomalies, Characters, And Strings,''
Nucl. Phys. {\bf B287} (1987) 317.}
\REF\witten{E. Witten, ``Elliptic Genera And Quantum Field THeory,''
Commun. Math. Phys. {\bf 109} (1987) 525, ``The Index Of The Dirac
Operator In Loop Space,'' in P. Landwebber, ed., {\it Elliptic Curves And
Modular Forms In Algebraic Topology} (Springer-Verlag, 1988).}
As we will see, an interesting and particularly rich
comparison  between Landau-Ginzburg and
minimal models can be made by consideration of the elliptic genus.
This is simply a genus one path integral with certain twisted boundary
conditions [\warner,\witten].  It has many convenient properties.
The elliptic genus of a supersymmetric model is conformally
invariant even if the underlying model is not conformally invariant.
There is, therefore, no difficulty in taking the infrared limit, which
simply coincides with the ultraviolet limit.  Moreover, the elliptic genus is
usually effectively computable, and this is so for both Landau-Ginzburg
and minimal models.  This is related to the fact that the elliptic
genus (of a sigma model, for instance) is a topological invariant
of the target space.

In \S2, we will compute the elliptic genus of the $N=2$ minimal models
and of the Landau-Ginzburg models introduced above.
By equating the formulas, we will obtain simple free field formulas
for certain characters of the $N=2$ algebra that follow from the
conjectured relation between these models.
We will verify the first few terms of these formulas.
A general verification of them has been proposed by N. Warner.

In \S3, we study the mechanism that underlies the character formula
proposed in \S2.  In fact, we will see that an $N=2 $ algebra, with
the expected central charge, and acting on a module with the character
found in \S2, can be constructed directly in the non-critical
Landau-Ginzburg model.  This leads naturally to a free field
realization of the $N=2$ algebra together with a construction
of a screening charge.
(We will not get a proof of the formula of \S2 since we will not prove
that the module is irreducible.)  The considerations of \S3 are a special
case of a general procedure for extracting a chiral algebra (in the sense
of rational conformal field theory) from any $N=2$ supersymmetric model
with $R$ symmetry.  The resulting chiral algebras seem worthy of further
study.

In sum, we will identify directly in the non-critical Landau-Ginzburg
model an $N=2$ algebra with the expected central charge and
a formula for certain of the $N=2$ characters.

This paper is dedicated to Professor  C. N. Yang on the occasion of
his 70th birthday.  Symmetries and interactions of elementary particles
have, of course,  always been foremost in his work.  He has also made lasting
contributions to statistical mechanics and many-body
physics, including
but not limited to the study of exactly soluble models in $1+1$ dimensions.
The $N=2$ superconformal
algebra in two dimensions is a relatively new symmetry structure, linked
on the one hand to soluble models in $1+1$ dimensions, and on the other hand
to novel constructions in string theory
of models of elementary particle physics.  So I hope that the modest
contribution to understanding this algebra that I make in the present
paper is appropriate on this occasion.

\chapter{The Character Formula}
\section{The Elliptic Genus Of The Minimal Models}

The characters of the $N=2$ superconformal algebra in two dimensions
are functions of two variables, since the algebra contains a $U(1)$
charge $J_0$ that commutes with the Hamiltonian $H=L_0$.
The supercurrents $S_{\pm}$ have eigenvalues $\pm 1$ under $J_0$, in the
sense that $[J_0,S_\pm]=\pm S_{\pm}$.
It follows that
$$\exp(i\pi J_0) S_\pm \exp(-i\pi J_0) =-S_\pm, \eqn\hido$$
while $\exp(i\pi J_0)$ commutes with the bosonic generators
of the $N=2$ algebra, namely the current $J$ and stress tensor $T$:
$$\exp(i\pi J_0)\left\{\matrix{ J\cr T\cr}\right\}\exp(-i\pi J_0)
 =\left\{\matrix{ J\cr T\cr}\right\}.  \eqn\jco$$

Relations \hido\ and \jco\ are also valid if $\exp(i\pi J_0)$ is
replaced by the operator $(-1)^F$ that is $+1$ for bosons and
$-1$ for fermions.  It follows that the product $(-1)^F\exp(i\pi J_0)$
is central, and so is a complex constant in any irreducible representation
of the $N=2$ algebra.

We will take the definition of the character of a representation $R$
of the $N=2$ algebra to be
$$\chi_R(q,\gamma)=\Tr_R(-1)^Fq^He^{i\gamma J_0} . \eqn\polyp$$
The factor of $(-1)^F$ could be replaced by $\exp(i\pi J_0)$ in view
of the above remarks.

Of course, in physical terms, the $N=2$ algebra, in its irreducible
highest weight representations, acts in a Hilbert space of only
left-moving (or only right-moving) degrees of freedom.
The $N=2$ minimal models are quantum field theories obtained
by combining left- and right-movers.
There are several ways to do this, with an ${\bf A}
-{\bf D}-{\bf E}$ classification;
for simplicity, we will consider only the ${\bf A}$ series.  This is
constructed
as follows: if $R_\alpha$ are the irreducible representations of
a left-moving $N=2$ algebra, and $\overline R_\alpha$ are the complex
conjugate representations of a right-moving $N=2$ algebra, then
the Hilbert space of the ${\bf A}$ series of $N=2$ minimal models
is ${\cal H}=\oplus_\alpha R_\alpha\otimes \overline R_\alpha$.
Nothing essential is lost if we consider only the case that both
left-movers and right-movers are in the Ramond sector; corresponding
results for the Neveu-Schwarz sector can be obtained by spectral flow.

Let $H_L$ and $H_R$ ($=\overline L_0,L_0$) be the Hamiltonians
of left-movers and right-movers; and let $J_{0,L}$ and  $ J_{0,R}$ be the
$U(1)$ charges of the left-movers and right-movers.
The natural generalization of \polyp\ to include both left- and right-movers
is
$$\eqalign{
Z(q,\gamma_L,\gamma_R)=&\Tr_{\cal H}(-1)^Fq^{H_L}\overline q^{H_R}
\exp(i\gamma_LJ_{0,L}+i\gamma_RJ_{0,R})  \cr = &
\sum_\alpha\Tr_{R_\alpha}(-1)^{F_L}q^{H_L}
\exp(i\gamma_LJ_{0,L})\,\,\Tr_{\overline R_\alpha}(-1)^{F_R}\overline q^{H_R}
\exp(i\gamma_RJ_{0,R}).\cr}  \eqn\imbob$$
Here we have factored $(-1)^F$ as
$(-1)^F=(-1)^{F_L}(-1)^{F_R}$, where $(-1)^{F_L}$ and $(-1)^{F_R}$ act
in left- and right-moving Hilbert spaces.

The partition function $Z(q,\gamma_L,\gamma_R)$
is effectively computable for the $N=2$ minimal
models by algebraic methods.  However, our goal is really to compare
the minimal models to Landau-Ginzburg models, and the full partition
function is not effectively computable
in the Landau-Ginzburg representation.  The situation changes markedly
if we consider not the full character, but the elliptic genus, which
is simply the partition function restricted to $\gamma_R=0$.  This
specialization of the partition function
is effectively computable in the Landau-Ginzburg
representation, as we will see later.  For the moment,
we work out the elliptic genus of the minimal models.  The right-moving
factor in \imbob\ is simply $\Tr_{\overline R_\alpha}(-1)^{F_R}\overline
q^{H_R}$.  As is usual in index theory,
because of bose-fermi cancellation, this expression can
be evaluated just by counting states of $H_R=0$.
It equals $+1$ for $\alpha$ such that the ground state of the
representation $R_\alpha$ has $H_R=0$, and 0 for other $\alpha$.
The elliptic genus is hence
$$Z(q,\gamma_L,0)=\sum\nolimits'_\alpha \Tr_{R_\alpha}(-1)^{F_L}q^{H_L}
\exp(i\gamma_LJ_{0,L})=\sum\nolimits'_\alpha \chi_\alpha(q,\gamma_L),
\eqn\concon$$
where $\sum'_\alpha$ is a sum restricted to $\alpha$ such that the
vacuum vector of $R_\alpha$ has $H_R=0$.  These are precisely the
representations which upon spectral flow to the Neveu-Schwarz sector
correspond to the chiral primary states.

{}From \concon, it may appear
that the elliptic genus determines not individual characters of the
$N=2 $ algebra, but only certain linear combinations of them.  However,
by using some information about the spectrum of $J_{0,L}$, it is possible
to invert this relation to express (certain) characters of the $N=2$ algebra
in terms of the elliptic genus.  To begin with, in the Ramond sector,
the $U(1)$ charges of the chiral primary states
are $n/(k+2)$, with $n=0,1,2,\dots, k$.  (These states are represented
in the Landau-Ginzburg language as $\Phi^n$, where $\Phi$ has charge
$1/(k+2)$.)  Under spectral flow, these states become Ramond sector
ground states with $J_{0,L}$ eigenvalue
$$q_n=-{\widehat c\over 2}+{n\over k+2},\eqn\ipool$$
where for these particular models the central charge is
$$\widehat c = 1-2\alpha \eqn\mippol$$
with
$$\alpha={1\over k+2}.\eqn\mcmmccx$$
Let $R_n$, $n=0,\dots ,k$ be the Ramond sector representation of the
$N=2$ algebra containing a ground state of $H=0$ and $J_{0}=q_n$.
Let
$$\chi_n(q,\gamma)=\Tr_{R_n}(-1)^Fq^H\exp(i\gamma J_0) \eqn\nurg$$
for $0\leq n\leq k$, and $\chi_{k+1}(q,\gamma)=0$.
The eigenvalues of $J_{0,L}$ in the representation $R_n$ are congruent
to $q_n$ modulo ${\bf Z}$, so
$${1\over k+2}\sum_{m=0}^{k+1}\chi_n(q,\gamma+2\pi m)\exp(\pi i\widehat
c m-2\pi i sm
\alpha)
=\delta_{s,n}\chi_n(q,\gamma).
\eqn\overflo$$
Upon summing over $n$ and using \concon, this gives
$$\chi_s(q,\gamma)={1\over k+2}\sum_{m=0}^{k+1}Z(q,\gamma+2\pi m,0)
 \exp\left({\pi i\widehat c m - 2\pi i\alpha ms}\right).
 \eqn\juncoc$$

\section{The Elliptic Genus Of A Landau-Ginzburg Model}

We now wish to compute the function $Z(q,\gamma,0)$ in the Landau-Ginzburg
model.  The first step is to identify the left- and right-moving
$U(1)$ charges.  The supersymmetry transformation laws of the
Landau-Ginzburg model, after eliminating the auxiliary field $F$
by its equation of motion, are (with conventions as in [\bagger,\lastwitten])
$$\eqalign{\delta\phi & = \sqrt 2\left(-\epsilon_-\psi_++\epsilon_+\psi_-
         \right) \cr
           \delta\psi_+ & = i\sqrt 2\left(\partial_0+\partial_1\right)\phi
\overline\epsilon_-+\sqrt 2\epsilon_+\overline\phi^{k+1} \cr
\delta\psi_- & = -i\sqrt 2
\left(\partial_0-\partial_1\right)\phi\overline\epsilon_+
+\sqrt 2\epsilon_-\overline\phi^{k+1}.\cr}\eqn\jucuc$$
The Lagrangian of the model, after integrating out fermionic coordinates
and eliminating the auxiliary field, is
$$\eqalign{L=\int\d^2x &\left(-\partial_\alpha\overline\phi\partial^\alpha\phi
+i\overline\psi_-(\partial_0+\partial_1)\psi_-+i\overline\psi_+
(\partial_0-\partial_1)\psi_+\right.\cr &\left.-(\overline\phi \phi)^{k+1}
-(k+1)\phi^k\psi_-\psi_+-(k+1)
\overline\phi^k\overline\psi_+\overline\psi_-\right)
.\cr}\eqn\hducu$$

The left-moving $U(1)$ charge $J_{0,L}$ generates a symmetry of the Lagrangian
\hducu\ under which the supersymmetry generators transform as follows: the
generator $\epsilon^+=-\epsilon_-$ (which generates a symmetry that in
the massless limit couples only to right-movers) should be invariant,
while $\epsilon^-=\epsilon_+$ should have charge 1.  These conditions
uniquely identify the symmetry generated by $J_{0,L}$ to be
$$\eqalign{
  \phi & \to \exp\left({i\gamma\alpha}\right)\phi \cr
  \psi_+ & \to \exp\left({i\gamma\alpha}\right)\psi_+ \cr
\psi_- & \to \exp\left(-{i(k+1)\gamma\alpha}\right)\psi_- .\cr}\eqn\jhdj$$
Thus, if the conjectured relation between the Landau-Ginzburg model
and the $N=2$ minimal model is correct, then the left-moving $U(1)$ symmetry
of the minimal model must correspond under this
relation to the symmetry group \jhdj.

\subsection{The Elliptic Genus}

We wish to consider path integrals
on a genus one Riemann surface,
\foot{We take this surface to be of positive signature.  The relation
to the Lorentz signature formulas used elsewhere in this paper is
by a standard Wick rotation $x^0=-ix^2$.}
say the torus $\Sigma$ obtained by dividing
the $x^1-x^2$ plane by the equivalence relation $x^1\to x^1+m,\,\,
x^2\to x^2+n$, with $m,n\in {\bf Z}$.  Given a particular physical
model, various path integrals can be defined on $\Sigma$, depending
on the boundary conditions that one chooses for the fields.  For example,
the simplest supersymmetric index $\Tr(-1)^F$ corresponds to using untwisted
boundary conditions for all bosons and fermions in both the $x^1$ and $x^2$
direction.  The elliptic genus is obtained, instead,
by twisting the fields by a left-moving $R$ symmetry, that is a symmetry
that commutes with the right-moving supersymmetries but not with the
left-moving ones.  In a model
with $N=1$ supersymmetry, the $R$ symmetry group (if any) is ${\bf Z}_2$,
and the usual elliptic genus is obtained by twisting by the non-trivial
element of ${\bf Z}_2$.  With $N=2$ supersymmetry, the $R$ symmetry
group can be $U(1)$ (as in the models considered in this paper), and
then one can twist by arbitrary elements of $U(1)$.

The elliptic genus $Z(q,\gamma,0)$ of the $N=2$ minimal
models was defined by the formula
$$Z(q,\gamma,0)=
\Tr_{{\cal H}}(-1)^F q^{H_L}\overline q^{H_R}\exp(i\gamma J_{0,L}).
\eqn\cccncn$$
Here ${\cal H}$ was an ordinary, untwisted Hilbert space, so the
path integral representation of this quantity involves untwisted boundary
conditions in the ``space'' direction, which we can take to be the $x^1$
direction.  The factor of $\exp(i\gamma J_{0,L})$ means that in the ``time''
direction, which we can take to be the $x^2$ direction, the fields
must be twisted by $\exp(i\gamma J_{0,L})$.
In the Landau-Ginzburg model, this means by virtue of \jhdj\
that the fields obey
$$\eqalign{
  \phi(x^1,x^2+1) & =\exp\left({i\gamma\alpha}\right)\phi(x^1,x^2) \cr
  \psi_+(x^1,x^2+1) & =\exp\left({i\gamma\alpha}\right)\psi_+(x^1,x^2) \cr
\psi_-(x^1,x^2+1)
 & = \exp\left(-{i(k+1)\gamma\alpha}\right)\psi_-(x^1,x^2) .\cr}
\eqn\jhdjj$$
And of course they are invariant under $x^1\to x^1+1$.

Can the Landau-Ginzburg path integral
with these boundary conditions be effectively evaluated?
Landau-Ginzburg path integrals are intractable in general, but
here the topological invariance of the elliptic genus comes to our aid.
The elliptic genus has an interpretation (which we recall at the beginning of
\S3) as an index of a right-moving
supercharge; this  ensures that it is invariant under continuous variations
of the Lagrangian of a supersymmetric system -- as long as one considers
only systems with a good behavior for large values of the fields, so
that low energy states cannot appear or disappear at infinity in field
space.  As a result, the elliptic genus would be unchanged if one
replaces the superpotential $W(\Phi)=\Phi^{k+2}/(k+2)$ by
$\widetilde W(\Phi)=\epsilon \Phi^{k+2}/(k+2)$ for any non-zero
$\epsilon$.  One is tempted to try to take the limit $\epsilon\to 0$
to get a free field theory.  This is dangerous because precisely at
$\epsilon=0$ the $\phi$ field can be arbitrarily large at no cost in energy.

For instance, the supersymmetric index $\Tr(-1)^F$ is not continuous
at $\epsilon=0$; it equals $k+1$ for $\epsilon\not= 0$, and is ill-defined
at $\epsilon=0$ because in finite volume the zero mode of the $\phi$ field
gives the Hamiltonian of the system a continuous spectrum going down to
zero energy. The path integral representation of $\Tr(-1)^F$ involves
untwisted boundary conditions for $\phi$ (and all other fields) in the
$x^1$ and $x^2$ directions, and is ill-behaved at $\epsilon=0$ because
one loses control over the zero mode of the $\phi$ field.

\REF\vvafa{C. Vafa, ``String Theory And Orbifoldized LG Models,''
Mod. Phys. Lett. {\bf A4} (1989) 1169.}
For the elliptic genus, the situation is completely different.
The twisted boundary conditions \jhdj\ remove the zero mode of $\phi$,
and ensure that the path integral is convergent even at $\epsilon=0$
-- and that there is no
problem in formal arguments showing that $Z(q,\gamma,0)$ is independent
of $\epsilon$.
(This generalizes the fact that supersymmetric ground states in
twisted sectors of Landau-Ginzburg models can be computed via free
field theory [\vvafa].)
So we can simply set $\epsilon=0$,
and we get a free field representation of the elliptic genus of the
Landau-Ginzburg model.

If the Landau-Ginzburg model has the conjectured
relation to the $N=2$ minimal model, this will give us a free field
representation of the elliptic genus of the minimal model and therefore
(according to \juncoc) of certain $N=2$ characters.

We could of course generalize the definition of the function $Z(q,\gamma,0)$
to allow for twists by the left-moving $U(1)$ symmetry in the $x^1$
as well as $x^2$ direction.  This has one theoretical advantage:
it eliminates the bosonic zero mode from the Hamiltonian as well as the
path integral formulation.

\subsection{Free Field Computation}

The free field computation of the twisted partition function that
gives the elliptic genus can perhaps be performed most conveniently
in a Hilbert space approach.

Let us work out the contributions
of the fermionic and bosonic zero and non-zero
modes to the elliptic genus defined as in  \cccncn.
We begin with the fermionic zero modes.  The zero modes of $\psi_-$
and $\overline \psi_-$ -- call them $\psi_{-,0}$ and $\overline\psi_{-,0}$ --
obey $\{\psi_{-,0},\overline\psi_{-,0}\}=1$.  So
 they are represented in a space of two states
$|\downarrow\rangle$ and $|\uparrow\rangle$, with
$$\psi_{-,0}|\downarrow\rangle
=|\uparrow\rangle,~~~\overline\psi_{-,0}|\uparrow\rangle=|\downarrow\rangle
\eqn\hudn$$
and other matrix elements zero.
In view of the quantum numbers of $\psi_{-,0}$ and $\overline\psi_{-,0}$,
one of the states $|\uparrow\rangle,\,\,\,|\downarrow\rangle$
is bosonic and one is fermionic; and they
transform under the $U(1)$ symmetry \jhdj\ as $\exp(\mp i\gamma(k+1)\alpha/2
)$.
The contribution of these two states to \cccncn\ is therefore
(up to an overall sign that can be absorbed in the definition of the operator
$(-1)^F$) a factor of
$$e^{-{i\gamma(k+1)\alpha/2}}-e^{{i\gamma(k+1)\alpha
/2}}.\eqn\jdjdxx$$
Similarly, the zero modes of $\psi_+,\overline\psi_+$ contribute a
factor of
$$e^{i\gamma\alpha/2}-e^{-i\gamma\alpha/2}
. \eqn\ruruuu$$
The overall factor coming from fermi zero modes is the product of these
factors;
this can be written
$$e^{-i\gamma k\alpha/2}\cdot\left(1-e^
{i\gamma(k+1)\alpha}\right)
\cdot \left(1-e^{-i\gamma\alpha}\right).
\eqn\zmodes$$

The non-zero modes of left- and right-moving fermions contribute
a factor
$$\prod_{n=1}^\infty \left(1-q^ne^{-i(k+1)\gamma\alpha}\right)
\left(1-q^ne^{i(k+1)\gamma\alpha}\right)\left(1-\overline q^ne^{i\gamma\alpha}
\right)\left(1-\overline q^ne^{-i\gamma\alpha}\right)\eqn\hdhcx$$
coming from a trace over the fermion Fock space.
The first two factors in \hdhcx\ come from left-movers and the
last two from right-movers.  The non-zero modes of bosons contribute
an analogous factor
$$\prod_{n=1}^\infty {1\over 1-q^ne^{i\gamma\alpha}}
\cdot {1\over 1-q^ne^{-i\gamma\alpha}}
\cdot {1\over 1-\overline q^ne^{i\gamma\alpha}}
\cdot {1\over 1-\overline q^ne^{-i\gamma\alpha}},\eqn\hdhcxx$$
from a trace over the bosonic Fock space.  The contribution of the bosonic
zero modes is actually given by a factor of the same structure, namely
$${1\over 1-e^{i\gamma\alpha}}\cdot {1\over 1-e^{-i\gamma\alpha}},\eqn\cxcx$$
even though the Hilbert space of the bosonic zero modes (which is naturally
represented as ${\bf L}^2({\bf C})$, where ${\bf C}$ is a copy of
the complex plane parametrized by the zero mode of the $\phi$ field)
has no natural representation as a Fock space.
One way to justify \cxcx\ is to perturb
the definition
of $Z(q,\gamma,0)$ to introduce a small twist by the left-moving
$U(1)$ charge in the $x^1$ direction.  This has the effect of displacing
the bosonic zero mode of the Hamiltonian description, giving it a non-zero
energy, so that the Hilbert space has a standard Fock space representation,
and the would-be zero mode can manifestly be treated in the same
way as the other bosonic modes.  Upon removing the twist in the $x^1$
direction,
one gets \cxcx\ as the zero mode contribution.  (Alternatively, one can
get \cxcx\ from the path integral, which because of the twist in the $x^2$
direction has no problem with zero modes.)

So putting the factors together, we find for the elliptic genus of the
Landau-Ginzburg model
$$\eqalign{Z(q,\gamma,0)=e^{-i\gamma k\alpha/2}\cdot
{1-e^{i\gamma(k+1)\alpha}\over
1-e^{i\gamma \alpha}}
\cdot\prod_{n=1}^\infty{(1-q^ne^{i\gamma(k+1)\alpha})(1-q^ne^{-i\gamma
(k+1)\alpha})\over
(1-q^ne^{i\gamma\alpha})(1-q^ne^{-i\gamma
\alpha})}.\cr}\eqn\finalformula$$
If Landau-Ginzburg models are related to $N=2$ minimal models in the
conjectured fashion, then via \juncoc, we get from this an explicit
formula for certain characters of the $N=2$ superconformal algebra.
Indeed, the Ramond sector character $\chi_s(q,\gamma)$ corresponding
to the chiral primary field $\Phi^s$ must be
$$\chi_s(q,\gamma)={1\over k+2}\sum_{m=0}^{k+1}Z(q,\gamma+2\pi m,0)
 \exp\left({i\pi \widehat c m - 2\pi ims\alpha}\right),
 \eqn\bbuncoc$$
with $Z(q,\gamma,0)$ as in \finalformula.

It is straightforward to expand the above formula for the $\chi_s$
in powers of $q$ and verify the first few terms.
For instance, verifying the formulas for the first two levels is
straightforward using the formulas
$$\eqalign{\chi_0 & = e^{-i\gamma k\alpha/2}\left(1+q(1-e^{i\gamma})
+O(q^2)\right) \cr
\chi_n & = e^{-i\gamma k\alpha/2}e^{i\gamma n \alpha}
\left(1+q(2-e^{i\gamma}-e^{-i\gamma})
+O(q^2)\right), ~~{\rm for}~0<n<k \cr
\chi_k & = e^{i\gamma k\alpha/2}\left(1+q(1-e^{-i\gamma})
+O(q^2)\right) \cr
\chi_{k+1} & = 0 \cr} \eqn\hducuc$$
which follow easily from the $N=2$ algebra.
Apart from the vanishing of $\chi_{k+1}$, which reflects the fact
that $\Phi^{k+1}$ is not a primary field, these formulas express
the fact that all the $\chi_s$ have highest weight states of zero energy,
and $\chi_0$ and $\chi_k$ have additional null vectors of level one,
because they are related
under spectral flow to the representation containing the identity operator.

\REF\dobrev{V. K. Dobrev, ``Characters Of The Unitarizable Highest
Weight Modules Over The $N=2$ Superconformal Algebra,'' Phys. Lett.
{\bf B186} (1987) 43.}
\REF\rav{F. Ravanini and S.-K. Yang, ``Modular Invariance In $N=2$
Superconformal Field Theory,'' Phys. Lett. {\bf B195} (1987) 202.}
\REF\qiuq{Z. Qiu, ``Modular Invariant Partition Functions For $N=2$
Superconformal Field Theory,'' Phys. Lett. {\bf B198} (1987) 497.}
\REF\dq{J. Distler and Z. Qiu, ``BRS Cohomology And A Feigin-Fuchs
Representation of Kac-Moody and Parafermionic Theories,'' Nucl. Phys.
{\bf B336} (1990) 533.}
\REF\eguchi{T. Eguchi, T. Kawai, S. Mizoguchi, and S.-K. Yang,
``Character Formulas For Coset $N=2$ Superconformal Theories,''
IEK Preprint 91-111.}
\REF\kedem{R. Kedem, T. R. Klassen, B. M. McCoy, and E. Melzer,
``Fermionic Sum Representations For Conformal Field Theory Characters,''
Stonybrook preprint ITP-SB-93-0.}
A general proof of the above formulas for $\chi_s$ has been proposed by
N. Warner (private communication), by comparing them to previously known
character formulas [\dobrev--\kedem].  I believe that it should
be possible to obtain a more conceptual proof along the following lines.
The ${\bf A}$ series of $N=2$ minimal models is known to have a representation
as a gauged WZW model of $SU(2)/U(1)$.  It should be possible
to compute the elliptic genus of a gauged WZW model along the same
lines as the above, and in the special case of $SU(2)/U(1)$, this is likely
to give back the above formulas.

In the rest of this paper, I will pursue a somewhat different
approach to better understanding of \bbuncoc.
I will extract from the Landau-Ginzburg model
a free field representation of the $N=2$ superconformal algebra,
in a module with character \bbuncoc.

\chapter{The Landau-Ginzburg Model And The $N=2$ Algebra}

\section{$N=1$ Index and $N=2$ Cohomology}

First we recall the interpretation of the elliptic genus as an index.
For this, it suffices to have global $N=1$ supersymmetry, with
a ${\bf Z}_2$ $R$ symmetry.  Such a symmetry
is a factorization
$$(-1)^F=(-1)^{F_L}(-1)^{F_R}, \eqn\jongo$$
where $(-1)^{F_R}$ anticommutes with the right-moving supercharge $Q_R$
and commutes with the left-moving supercharge $Q_L$, and vice-versa
for $(-1)^{F_L}$.  In string theory, such a factorization is used in performing
the GSO projection.

The right-moving part of the $N=1$
supersymmetry algebra is $Q_R{}^2=H_R$, where $H_R=L_0=(H+P)/2$, with $H$
and $P$ being the Hamiltonian and the momentum.
The index of $Q_R$, regarded as
an operator from states of $(-1)^{F_R}=1$ to states of $(-1)^{F_R}=-1$, is
by a standard argument
the difference between the number of bosonic and fermionic states of
$H_+=0$, assuming these numbers are finite.
One can write it as
$$\Tr(-1)^{F_R}\overline q^{H_R}, \eqn\jcjc$$
if that trace converges.
Standard arguments show that states of $H_R\not= 0$ cancel out in pairs
in \jcjc, so if the spectrum is such that \jcjc\ is convergent,
this expression is independent of $\overline q$ and is a topological
invariant.

The index of $Q_R$ in that naive sense is not
defined in most quantum field theories, because we have included no
convergence factor for the left-movers.
In general, it is essential to consider a twisted
or character-valued version of the index, a quantity such as the elliptic
genus
$$ \Tr(-1)^{F_R}\widetilde q^{H_R}
q^{H_L}\cdot X\eqn\kdkkd$$
with $|q|,|\widetilde q|<1$, $H_L=\overline L_0$, and $X$ any operator
that commutes with $Q_R$.  (In $N=2$ models with continuous $R$ symmetry,
it is convenient to take $X=(-1)^{F_L}\exp(i\gamma J_{0,L})$ as in \S2.)
Using the fact that $q^{H_L}X$ commutes with
$Q_R$, one can show that states of $H_R\not= 0$ cancel out in pairs from this
trace.  Hence if (as in most interesting field theories), the
$H_L$ eigenvalues of states of $H_R=0$ grow fast enough that the trace
in \kdkkd\ converges, this trace
is independent of $\widetilde q$ (and so is holomorphic
in $q$ if, as usual, one sets $\widetilde q=\overline q$).

So far, we have used only
$N=1$ supersymmetry and discrete $R$ invariance.  With $N=2$ supersymmetry,
there are two right-moving supercharges, $Q_+$ and $\overline Q_+$,
with $Q_+{}^2=\overline Q_+{}^2=0$, $\{Q_+,\overline Q_+\}=2H_R$.
The fundamental novelty is that, using the fact that $\overline Q_+{}^2=0$,
one can define the {\it cohomology} of $\overline Q_+$.  By standard
arguments of Hodge theory, the cohomology of $\overline Q_+$ can be identified
with the kernel of $H_R$.  Hence, any trace such as \kdkkd\ can be regarded
as a trace in the cohomology of $\overline Q_+$.  However, by considering
the cohomology of $\overline Q_+$ one has much more structure than if one
considers only the graded traces in this cohomology.  The cohomology is a
graded vector space, and one can perform operations in this space other
than taking cohomology.

\def\oq{\overline Q_+}
For instance, one can look for operators that commute
with $\overline Q_+$ and so act on the cohomology of this operator.
Operators of the form $\{\oq,\dots\}$ will act trivially on the cohomology
of $\oq$, so the natural problem is to consider operators that commute
with $\oq$ modulo operators of the form $\{\oq,\dots\}$; that is, we are
interested in cohomology classes of operators that commute with $\oq$.
If $\{\oq, {\cal O}\}=\{\oq, {\cal O}'\}=0$, then $\{\oq,{\cal O}{\cal O}'\}
=0$, and if $\{\oq, {\cal O}\}=0$, then ${\cal O}\{\oq, X\}=
\{\oq, {\cal O}X\}$.  These two statements mean that the cohomology classes
of operators that commute with $\oq$ form a closed (and well-defined)
algebra under operator products.

\REF\brivafa{C. Vafa, ``Superstring Vacua,'' in the proceedings of the
Trieste School on High Energy Physics, 1989.}
Moreover, if $\{\oq,{\cal O}\}=0$, then $[H_R,{\cal O}]=\{\oq,\{Q_+,{\cal O}\}
\}/2$,
or more briefly $[H_R,{\cal O}]=\{\overline Q_+,\dots\}$.  Hence, operators
invariant under $\oq$ commute with $H_R$ up to $\{\oq,\dots\}$.  Thus,
$H_R$ annihilates cohomology classes of $\oq$-invariant operators.
An operator annihilated by $H_R$ varies holomorphically on the world-sheet.
All this can be summarized by saying
that the cohomology of $\oq$ has the structure of
a closed algebra of operators that vary holomorphically.  This structure
is a (not necessarily unitary) chiral algebra in the sense of rational
conformal field theory.  This has been noted earlier (see question 7
in [\brivafa]).

Along with WZW models and their derivatives -- which have been extensively
studied -- this construction seems to give one of the few known sources
of such chiral algebras.

In what follows, we will study the chiral algebras
acting on the cohomology
of an $N=2$ Landau-Ginzburg
model with one chiral superfield $\Phi$ and superpotential $W(\Phi)
=\Phi^{k+2}/(k+2)$.  We will show that the chiral algebra derived
from this theory is an $N=2$ superconformal algebra with the
central charge of the $N=2 $ minimal models.  Since this algebra acts
naturally in the cohomology of $\oq$, which is a graded vector space
whose character is the candidate \bbuncoc\ for characters of
the $N=2$ algebra, this gives a natural explanation of \bbuncoc\
(but not yet a proof as we will not prove irreducibility of the action of
the $N=2$ algebra on the cohomology).

\REF\zam{A. B. Zamolodchikov and V. A. Fateev, ``Disorder Fields In
Two-Dimensional Conformal Quantum-Field Theory And $N=2$ Extended
Supersymmetry,'' Sov. Phys. JETP {\bf 63} (1986) 913.}
\REF\nam{S. Nam, ``The Kac Formula For The $N=1$ and $N=2$
Superconformal Algebras,'' Phys. Lett. {\bf B172} (1986) 323.}
\REF\mussardo{G. Mussardo, ``Fusion Rules, Four Point Functions, and
Discrete Symmetries Of $N=2$ Superconformal Models,'' Phys. Lett. {\bf B128}
(1989) 191, ``$N=2$ Superconformal Minimal Models,'' Int. J. Mod. Phys.
{\bf A4} (1989) 1135.}
\REF\ohta{N. Ohta and H. Suzuki, ``$N=2$ Superconformal Models
And Their Free Field Realizations,'' Nucl. Phys. {\bf B332} (1990) 146.}
\REF\ito{K. Ito, ``Quantum Hamiltonian Reduction And $N=2$
Coset Models,'' Phys. Lett. {\bf B259} (1991) 73.}
\REF\nem{D. Nemeschansky and S. Yankielowicz, ``$N=2$ $W$-algebras,
Kazama-Suzuki Models, and Drinfeld-Sokolov Reduction,'' USC preprint
USC-007-91 (1991).}
\REF\sadov{V. Sadov, ``Free Field Resolution For Nonunitary
Representations Of $N=2$ SuperVirasoro,'' Harvard preprint
HUTP-92/A070.}
\REF\fre{P. Fr\'e, F. Gliozzi, M. Monteiro, and A. Piras,
``A Moduli-Dependent Lagrangian For $(2,2)$ Theories On
Calabi-Yau $n$-Folds,'' Class. Quant. Grav. {\bf 8} (1991) 1455;
P. Fr\'e, L. Girardello, A. Lerda, and P. Soriani,
``Topological First-Order Systems With Landau-Ginzburg Interactions,''
Nucl. Phys. {\bf B387} (1992) 333.}
There are several notable features about the $N=2$ algebra that acts
on the cohomology of the Landau-Ginzburg model.
(1) This is a {\it superconformal} algebra even
though the underlying Landau-Ginzburg
model had global supersymmetry only.  Hopefully, its
occurrence is a harbinger of the conjectured
Landau-Ginzburg/minimal model connection.
(2)  Once the $N=2$ algebra
that acts on the cohomology is found, its definition continues
to make sense if the superpotential of the $N=2$ model
is set to zero.  We get then a {\it free field
realization} of the $N=2$ superconformal algebra; such representations
are known [\newv,\dq,\zam--\fre].  The one we obtain is not the one
that has been most frequently seen, but it can be obtained from that
of [\fre] by a simple transformation (regarding $\phi$ and $\partial
\bar\phi$ as independent variables $\beta$ and $\gamma$).  The
considerations of [\fre] are based on a simple model that in contrast
to the Landau-Ginzburg model has
a manifest $N=2$ superconformal symmetry (classically and quantum
mechanically) but no manifest properties of unitarity.
(3) The ``screening charge'' associated with this free field representation
(which was also introduced in [\fre])
can be simply identified by further study of $\oq$, making this one of
the few cases in which such a screening charge has a simple conceptual
explanation.

\section{The Free Field Realization}

We now want to find cohomology classes of the $\overline Q_+$ operator
of the Landau-Ginzburg model.  It is convenient to first transform
the problem as follows.

We recall that supersymmetry is realized in $N=2$ superspace by the
differential operators
$$\eqalign{ Q_\alpha & = {\partial\over\partial\theta^\alpha}-i
        \sigma^m_{\alpha\dot\alpha}\overline\theta{}^{\dot\alpha}
                     {\partial\over\partial x^m} \cr
     \overline Q_{\dot\alpha} & = -{\partial\over\partial \overline
            \theta{}^{\dot\alpha}}+i\sigma^m_{\alpha\dot\alpha}\theta^\alpha
                 {\partial\over\partial x^m} .\cr}\eqn\gimdo$$
These operators commute with the operators
$$\eqalign{ D_\alpha & = {\partial\over\partial\theta^\alpha}+i
        \sigma^m_{\alpha\dot\alpha}\overline\theta{}^{\dot\alpha}
                     {\partial\over\partial x^m} \cr
     \overline D_{\dot\alpha} & = -{\partial\over\partial \overline
            \theta{}^{\dot\alpha}}-i\sigma^m_{\alpha\dot\alpha}\theta^\alpha
                 {\partial\over\partial x^m} .\cr}\eqn\gimdox$$
The $Q$'s and $D$'s are related by formulas such as
$$\overline Q_{\dot\alpha}
= \exp\left(-2i\sigma^m_{\alpha\dot\alpha}\theta^\alpha
 \overline\theta^{\dot\alpha}{\partial\over\partial x^m}\right)
\overline D_{\dot\alpha}
\exp\left(2i\sigma^m_{\alpha\dot\alpha}\theta^\alpha
 \overline\theta^{\dot\alpha}{\partial\over\partial x^m}\right).
\eqn\jupol$$
Setting $\dot\alpha=+$, this formula means that the operators
$\overline Q_+$ and $\overline D_+$ are conjugate; hence instead
of computing the cohomology of $\overline Q_+$, we can compute the
cohomology of $\overline D_+$.  This is a more convenient formulation.

The superspace equations of motion derived from the superspace
Landau-Ginzburg Lagrangian \cambo\ are
$$ 2\overline D_+\overline D_- \overline \Phi -\Phi^{k+1} = 0 . \eqn\mxmx$$
A short computation shows that if
$${\cal J}={1\over 2}(1-\alpha)D_-\Phi\overline D_-\overline \Phi
-i\alpha \Phi(\partial_0-\partial_1)\overline\Phi, \eqn\uxuxu$$
then $\overline D_+{\cal J}=0$, and ${\cal J}$ cannot be written
as $\{\overline D_+,\dots\}$.
${\cal J}$ can be expanded in components to give various operators
that all represent non-trivial $\overline D_+$ cohomology classes:
$$\eqalign{
\left.{\cal J}\right|_{\theta=\overline\theta=0} & = (1-\alpha)\psi_-
\overline\psi_- -i\alpha\phi(\partial_0-\partial_1)\overline\phi \cr
\left.D_-{\cal J}\right|_{\theta=\overline\theta=0}& =-i\sqrt 2\psi_-
(\partial_0-\partial_1)\overline\phi  \cr
\left.\overline D_-{\cal J}\right|_{\theta=\overline\theta=0}&=
i\sqrt 2(1-\alpha)(\partial_0-\partial_1)\phi\cdot\overline\psi_-
-i\sqrt 2 \alpha\phi(\partial_0-\partial_1)\overline\psi_-   \cr
\left.{1\over 2}(\overline D_-D_--D_-\overline D_-)
{\cal J}\right|_{\theta=\overline\theta=0}&= 2(\partial_0-\partial_1)\phi\cdot
(\partial_0-\partial_1)\overline \phi+i\left(\psi_-(\partial_0-\partial_1)
\overline\psi_--(\partial_0-\partial_1)\psi_-\cdot\overline\psi_-\right)
\cr &~~~~~~+\alpha(\partial_0-\partial_1)\left(i\psi_-\overline \psi_--\phi
(\partial_0-\partial_1)\overline\phi\right).  \cr}
\eqn\ncncqq$$
As we will see, these components are respectively the $U(1)$ current,
the two supercurrents, and the stress tensor of an $N=2$ superconformal
algebra with $\widehat c = 1-2\alpha =1-2/(k+2)$.

We will verify this by computing the singular terms in the operator
products ${\cal J}(x,\theta){\cal J}(x',\theta')$.  In doing so,
we can ignore the part of the Lagrangian coming from the superpotential
because these terms are too ``soft'' to affect the singularities
of the ${\cal J}{\cal J}$ operator product.  Therefore, we can study
the operator products of the ${\cal J}$'s in free field theory.
Thus, our computation will amount to a demonstration that
the formulas \ncncqq\
give a free field representation of the $N=2$ algebra at level $k$.

According to [\bagger, eqn. 9.14], the free field propagator of
the superfield $\Phi$ is
$$\langle\Phi(x,\theta)\overline\Phi(x',\theta')\rangle
= -{1\over 4\pi}\ln(\widetilde x^m\widetilde x_m), \eqn\urmo$$
where
$$\widetilde x^m=(x-x')^m+i\theta\sigma^m\overline\theta+i\theta'\sigma^m
\overline\theta'-2i\theta\sigma^m\overline\theta'. \eqn\burmo$$
Using this propagator, it is straightforward but slightly lengthy
to compute the singular part of the operator products,
$$\eqalign{(4\pi)^2{\cal J}(x,\theta){\cal J}(0,0) &\sim
-{8\theta^-\overline\theta{}^-\over(x^0-x^1)^2}{\cal J}
-{2i\theta^-\over x^0-x^1}D_-{\cal J}-{2i\overline \theta^-\over x^0-x^1}
\overline D_-{\cal J}\cr &-{4\theta^-\overline \theta{}^-\over x^0-x^1}
(\partial_0-\partial_1){\cal J}-{4(1-2\alpha)\over (x^0-x^1)^2}+\dots
.\cr}             \eqn\cnnxo$$
This is a closed operator algebra, as expected; indeed, it is
an $N=2$ superconformal algebra with the expected central
charge $\widehat c=1-2\alpha$ (for instance, compare to equation
(10) of [\divec]).

I will not try to prove that the operator algebra generated
by ${\cal J}$ is the {\it full} chiral algebra acting on the cohomology
of the $\overline Q_+$ operator in the Landau-Ginzburg model.
However, this would be a consequence of our other conjecture,
which was that the $N=2$ algebra generated by ${\cal J}$ acts
irreducibly on the $\overline Q_+$ cohomology.

\subsection{Many Superfields}

Though we have focussed in this paper on Landau-Ginzburg models
with just one chiral superfield, most of our considerations
carry over to more general models.  Consider a Landau-Ginzburg
model with several chiral superfields $\Phi_i,\,\,\,i=1\dots n$
and superspace Lagrangian
$$L=\int \d^2x\,\,\d^4\theta\sum_i\overline\Phi_i\Phi_i
-\int\d^2x\,\d^2\theta \,\,W(\Phi_i)-\int \d^2x\,\d^2\overline \theta
\,\,\overline W(\overline\Phi_i). \eqn\conco$$
The equations of motion are
$$ 2\overline D_+\overline D_-\overline \Phi_i={\partial W\over\partial
\Phi_i}. \eqn\mcocnonc$$
The superpotential is said to be quasi-homogeneous if for some real
numbers $\alpha_i$, the Euler equation
$$ W=\sum_i\alpha_i\Phi_i{\partial W\over\partial\Phi_i}\eqn\xnxonxon$$
is obeyed.
If $W$ is quasi-homogeneous, a small calculation shows
that
$${\cal J}=\sum_i\left({1-\alpha_i\over 2}
D_-\Phi_i\overline D_-\overline\Phi_i
-i\alpha_i\Phi_i(\partial_0-\partial_1)\overline\Phi_i\right)
\eqn\xnonon$$
obeys $\overline D_+{\cal J}=0$.

In computing the singular part of the ${\cal J}{\cal J}$
operator products, the superpotential can be dropped, just as in the
one variable case.  The $\Phi_i$ can hence be treated as decoupled
free fields.  Formula \cnnxo\ from the case of one superfield
therefore carries over at once to show that ${\cal J}$ generates
an $N=2$ superconformal algebra; but the central charge receives
contributions from each of the $\Phi_i$ and so is now
$$\widehat c =\sum_i(1-2\alpha_i).      \eqn\coxp$$
This agrees with a formula of [\martinec--\lerche]
for the central charge
of the conformal field theory arising in the infrared from
a Landau-Ginzburg model with quasi-homogeneous superpotential.

One important difference from the one superfield case
is that in general one cannot expect ${\cal J}$ to generate
the full chiral algebra acting on the $\overline Q_+$
cohomology, only a sub-algebra.
It would be interesting to learn more about the full chiral algebra
that acts on the cohomology of a Landau-Ginzburg model.

\section{The Screening Charge}

The $\overline Q_+$ operator can be
written explicitly in components
$$\oq=\int \d x^1\left(i\overline\psi_+(\partial_0+\partial_1)\phi
+\phi^{k+1}\psi_-\right). \eqn\bimbo$$
We can write this $\oq=\overline Q_{+,L}+\overline Q_{_+,R}$,
where
$$\eqalign{
\overline Q_{+,R} & = \int \d x^1 \,\,i\overline \psi_+(\partial_0+
       \partial_1)\phi \cr
\overline Q_{+,L} & = \int \d x^1 \,\,\phi^{k+1}\psi_-.\cr} \eqn\bico$$
These obey $\overline Q_{+,R}{}^2=\overline Q_{+,L}{}^2=\{\overline Q_{+,R},
\overline Q_{+,L}\}=0$.

Because the Landau-Ginzburg theory is superrenormalizable and in fact
entirely free of divergences, the Hilbert space ${\cal H}$ of the interacting
theory can be taken to coincide with the Hilbert space of the free
theory with vanishing superpotential.
So it is natural to consider $\oq$ as an operator on the free field Fock space.
In so doing, one must specify how to treat the zero modes $\phi_0,\overline
\phi_0$
of $\phi,\overline\phi$; we will simply work in a Fock space of states
$|n,m\rangle =\phi_0{}^n\overline \phi_0{}^m|\Omega\rangle$.
\foot{This Fock space is precisely what one would have if one
``twists'' the theory slightly in the $x^1$ direction by $J_{0,L}$,
thus shifting the zero modes away from zero and reducing everything
to a standard Fock space.
This is a convenient artifice; however,  in a more natural formulation
of the problem,
one permits the zero mode part of the quantum wave-function to be
an arbitrary ${\bf L}^2$ function $\chi(\phi_0)$, with no natural
description as a Fock space.  In this case, it is still
true that the cohomology of $\oq$ coincides with the cohomology
of the operator $\overline Q_{+,L}$ acting on the space ${\cal H}_L$
defined in the text; but the argument required to prove this is a little
more elaborate.  The argument can be made by a very simple case of
a spectral sequence, filtering the Hilbert space ${\cal H}$ by an operator
that commutes with $\overline Q_{+,R}$ and under which $\overline Q_{+,L}$
has positive degree; then one computes the cohomology of $\oq$ by computing
first the cohomology of $\overline Q_{+,R}$.}

Now let ${\cal H}_R$ to be a Fock space of right-moving modes
of $\phi,\overline \phi, \psi_+$, and $\overline \psi_+$, and zero modes
of $\overline \phi, \psi_+,\overline \psi_+$;
and let ${\cal H}_L$ to be a Fock space of left-moving modes of
$\phi,\overline\phi,\psi_-$, and $\overline \psi_-$, and zero modes of
$\phi,\psi_-,\overline\psi_-$.
Then ${\cal H}={\cal H}_L\otimes {\cal H}_R$.  Moreover, $\overline Q_L$
acts in ${\cal H}_L$, and $\overline Q_R$ acts in ${\cal H}_R$.
So the cohomology of $\oq$ acting on ${\cal H}$
is simply the tensor product of the cohomology of $\overline Q_{+,L}$
acting on ${\cal H}_L$ and that of $\overline Q_{+,R}$ acting on
${\cal H}_R$.

But the cohomology of $\overline Q_{+,R}$ acting on ${\cal H}_R$
is one dimensional.  Indeed, as $\overline Q_{+,R}$ is quadratic
in the fields, its cohomology can be conveniently computed by going
to a basis of Fourier modes.  The part of $\overline Q_{+,R}$
involving the $n^{th}$ Fourier mode of $\phi,\psi_+$ (and so the $-n^{th}$
mode of $\overline \phi, \,\overline \psi_+$) is of the form
$$S_n=\left(\matrix{ 0 & A \cr 0 & 0 \cr}\right), \eqn\donothing$$
where $A$ is a bosonic creation or annihilation operator.  The cohomology
of $S_n$ is one dimensional, and $\overline Q_{+,R}$ is simply
an infinite sum of operators of this form, each acting on its own Hilbert
space;
and so $\overline Q_R$ has one
dimensional cohomology.

Hence the cohomology of $\oq$ is the same as the cohomology of $\overline
Q_{+,L}$ acting on ${\cal H}_L$.

Now, just like $\overline Q_{+,L}$,
the $N=2$ generator ${\cal J}$ acts in ${\cal H}_L$
(the only non-trivial point is that the $\phi$ zero mode but not the
$\overline\phi$ zero mode enters in the formula for ${\cal J}$).
The fact that ${\cal J}$ commutes with $\oq$ therefore means
that ${\cal J}$ commutes with $\overline Q_{+,L}$.  Hence, the $N=2$ algebra
generated by ${\cal J}$ actually acts on the cohomology of $\overline
Q_{+,L}$.  The Landau-Ginzburg/minimal model correspondence strongly
suggests that this action is unitary and irreducible.

In the language usually used in describing free field representations
of chiral algebras, the operator  $\overline Q_{+,L}$
is the ``screening charge'' of the free field representation of the
$N=2$ algebra that we have extracted from the Landau-Ginzburg model.
Like the construction of ${\cal J}$, the construction of $\overline
Q_{+,L}$ can be generalized directly to Landau-Ginzburg theories
with several superfields $\Phi_i$ and an arbitrary
superpotential $W$ (which in this case need not even be quasi-homogeneous).
The general formula is simply
$$\overline Q_{+,L}  = \int \d x^1 \,\,{\partial W\over\partial\phi_i}
\psi_{-,i}.\eqn\bicob$$
The chiral algebra of the $N=2$ model, which at a fundamental level
consists of fields that commute with $\oq$, is equivalent to the algebra
of local functions of left-moving free fields
that commute with the screening charge \bicob.

\endpage
\refout
\end